\pdfoutput=1
\documentclass[10pt,final,finalsubmission,twocolumn]{IEEEtran}

\usepackage{times}    
\usepackage{url}      
\usepackage{amssymb}
\usepackage{fancybox, graphicx}
\usepackage{algorithm}
\usepackage{algpseudocode}
\usepackage{subfigure}
\usepackage{slashbox}
\usepackage{epstopdf}
\usepackage{multirow}

\usepackage{color}

\graphicspath{{figures/}}

\makeatletter
\def\url@leostyle{%
  \@ifundefined{selectfont}{\def\UrlFont{\sf}}{\def\UrlFont{\small\bf\ttfamily}}}
\makeatother
\usepackage[pdftex]{hyperref}
\hypersetup{
pdftitle={SIGCHI Conference Proceedings Format},
pdfauthor={LaTeX},
pdfkeywords={SIGCHI, proceedings, archival format},
bookmarksnumbered,
pdfstartview={FitH},
colorlinks,
citecolor=black,
filecolor=black,
linkcolor=black,
urlcolor=black,
breaklinks=true,
}

\makeatletter

\newcommand{\Rmnum}[1]{\expandafter\@slowromancap\romannumeral #1@}
\makeatother

\begin{document}

\title{Blind Recognition of Touched Keys: Attack and Countermeasures}

\author{\IEEEauthorblockN{Qinggang Yue\IEEEauthorrefmark{1}, Zhen Ling\IEEEauthorrefmark{2}, Benyuan Liu\IEEEauthorrefmark{1}, Xinwen Fu\IEEEauthorrefmark{1} and Wei Zhao \IEEEauthorrefmark{3}\\
\IEEEauthorblockA{\IEEEauthorrefmark{1}University of Massachusetts Lowell, USA,
Email: \{qye,bliu,xinwenfu\}@cs.uml.edu} \\
\IEEEauthorblockA{\IEEEauthorrefmark{2}Southeast University, China,
Email: zhenling@seu.edu.cn} \\
\IEEEauthorblockA{\IEEEauthorrefmark{3}University of Macau, China,
Email: weizhao@umac.mo}
}}

\maketitle

%

\begin{abstract}
In this paper, we introduce a novel computer vision based attack that discloses inputs on a touch enabled device, while the attacker cannot see any text or popups from a video of the victim tapping on the touch screen. In the attack, we use the optical flow algorithm to identify touching frames where the finger touches the screen surface. We innovatively use intersections of detected edges of the touch screen to derive the homography matrix mapping the touch screen surface in video frames to a reference image of the virtual keyboard. 
We analyze the shadow formation around the fingertip and use the k-means clustering algorithm to identify touched points. Homography can then map these touched points to keys of the virtual keyboard.
Our work is substantially different from existing work. We target password input and are able to achieve a high success rate. We target scenarios like classrooms, conferences and similar gathering places and use a webcam or smartphone camera. In these scenes, single-lens reflex (SLR) cameras and high-end camcorders used in related work will appear suspicious. To defeat such computer vision based attacks, we design, implement and evaluate the Privacy Enhancing Keyboard (PEK) where a randomized virtual keyboard is used to input sensitive information.
\end{abstract}

\section{Introduction} \label{introduction}

Touch screen devices have been widely used since their inception in the 1970s. According to Displaybank's forecast
\cite{touchfuture}, 800 million smartphones are expected to be enabled by touch screen in 2014. Today, tablets, laptops, and ATM machines use touch screens, and touch-enabled devices have become part of our daily life. People use these devices to check bank accounts, send and receive emails, and perform various other tasks. Extensive private and sensitive information is stored on these devices.

Given their ubiquitous use in our daily life, touch enabled devices are attracting the attention of attackers. In March 2013, Juniper Networks reported that their Mobile Threat Center had discovered over 276 thousand malware samples, 614 percent increase over 2012 \cite{JuniperMTR::2013}. In addition to the threat of malware, one class of threats are computer vision based attacks. We can classify those attacks into three groups: the first group tries to directly identify the text on screen or its relfections on objects \cite{backes08compromising, backes09tempest}. The second group recognizes visible features of the keys such as light diffusion surrounding pressed keys \cite{clearShot} and popups of pressed keys \cite{eavesSmartphone, ispy}. The third group  blindly recognizes the text without visible text or popups. For example, Xu {\em et al}. track the finger movement to recover the input \cite{XHW+::SeeingDouble::2013}.

In this paper, we introduce an attack blindly recognizing input on touch enabled devices by recognizing touched points from a video of people tapping on the touch screen. Planar homography is employed to map these touched points to an image of virtual keyboard in order to recognize touched keys. Our work is substantially different from the work by Xu {\em et al}. \cite{XHW+::SeeingDouble::2013}, the most related work. First, we target password input while \cite{XHW+::SeeingDouble::2013} focuses on meaningful text so that they can use a language model to correct their prediction. In terms of recognizing passwords, we can achieve a high success rate in comparison with \cite{XHW+::SeeingDouble::2013}. Second, we employ a completely different set of computer vision techniques to track the finger movement and identify touched points more accurately to achieve a high success rate of recognizing passwords. Third, the threat model and attack scenes are different since targeted scenes are different. We study the privacy leak in scenes such as classrooms, conferences and other similar gathering places. In many such scenes, it is suspicious to face single-lens reflex (SLR) cameras with big lens and high-end camcorder with high optical zoom (used in \cite{XHW+::SeeingDouble::2013}) toward people. Instead, we use a webcam or smartphone camera for stealthy attack.

Our major contributions are two-fold. First, we introduce a novel computer vision based attack blindly recognizing touch input by recognizing and mapping touched points on the touch screen surface to a reference image of virtual keyboard. In the attack, an adversary first takes a video of people tapping from some distance, and preprocesses the video to get the region of interest, such as the touch screen area. The KLT algorithm \cite{KLT} is then used to track sparse feature points and an optical flow based strategy is applied to detect frames in which the finger touches the screen surface. Such frames are called {\em touching frames}.
We then derive the homography matrix between the touch screen surface in video frames and the reference image of the virtual keyboard. We innovatively use intersections of detected edges of the touch screen to derive the homography relation where SIFT \cite{siftFeature} and other feature detectors do not work in our context. We design a {\em clustering-based matching} strategy to identify touched points, which are then mapped to the reference image via homography in order to derive touched keys. We performed extensive experiments on iPad, Nexus 7 and iPhone 5 using a webcam or a phone camera from different distances and angles and can achieve a success rate of more than 90\%.

Our second major contribution is to design context aware randomized virtual keyboards, denoted as Privacy Enhancing Keyboard (PEK), to defeat various attacks. Intuitively, if keys on a virtual keyboard are randomized, most attacks discussed above will not work effectively. Of course, randomized keys incur longer input time. Therefore, our lightweight solution uses such a randomized keyboard only when users input sensitive information such as passwords. We have implemented two versions of PEK for Android systems: one using shuffled keys and the other with keys moving with a Brownian motion pattern. We recruit 20 people to evaluate PEK's usability. Our experiments show that PEK increases the password input time, but is acceptable for the sake of security and privacy to the interviewed people.

The rest of the paper is organized as follows: Section \ref{sec::RelatedWork} discusses most related work. Section \ref{problemDef} introduces the attack. We dedicate Section \ref{sec::KeyMatching} to discussing how to recognize touched points from touching images. Experimental design and evaluations are given in Section \ref{evaluation}. Section \ref{countermeasures} introduces PEK and its evaluation. We conclude this paper in Section \ref{conclusion}.

\section{Related work}
\label{sec::RelatedWork}

In this paper, we exploit the movement of the touching finger to infer the input on a touch screen. It is one kind of side channel attack. There are various such attacks on touch-enabled devices. Marquardt {\em et al}. use iPhone to sense vibrations of nearby keyboard \cite{iphoneVibration} to infer typed keys. Kune {\em et al}. \cite{timingAttack} collect and quantify the audio feedback to the user for each pressed button, use a Hidden Markov Model to narrow down the possible key space, and derive the keys. Aviv {\em et al}.\cite{smudgeSmartphone} expose typed keys by taking photos of the oil residue on touch screen while Zhang {\em et al}. \cite{fingerprintattack} apply fingerprint powder to a touch screen in order to expose touched keys.
Zalewski \cite{thermal} uses a thermal imaging camera to measure thermal residue left on a touched key to infer the touched key sequence. Mowery {\em et al}. perform a full scale analysis of this attack in \cite{thermalAnalysis}. Sensors including orientation sensor, accelerometer and motion sensors are also exploited to infer touched keys by correlating touching dynamics and key positions on touch screen \cite{Hao::TouchLogger::HotSec2011, OHDPZ::ACCessory::2012, XBZ::TapLogger::2012}.

In the following, we discuss the most related work on side channels using computer vision knowledge. Backes {\em et al}. \cite{backes08compromising, backes09tempest} exploit the reflections of a computer monitor on glasses, tea pots, spoons, plastic bottles, and eyes of the user to recover what is displayed on the computer monitor. Their tools include a SLR digital camera Canon EOS 400D, a refractor telescope and a Newtonian reflector telescope, which can successfully spy from 30 meters away.

Balzarotti {\em et al}. propose an attack retrieving text typed on a physical keyboard from a video of the typing process \cite{clearShot}. When keys are pressed on a physical keyboard, the light diffusing surrounding the key's area changes. Contour analysis is able to to detect such a key press. They employ a language model to remove noise. They assume the camera can see fingers typing on the physical keyboard.

Maggi {\em et al}. \cite{eavesSmartphone} implement an automatic shoulder-surfing attack against touch-enabled mobile devices. The attacker employs a camera to record the victim tapping on a touch screen. Then the stream of images are processed frame by frame to detect the touch screen, rectify and magnify the screen images, and ultimately identify the popping up keys.

Raguram {\em et al}. exploit refections of a device's screen on a victim's glasses or other objects to automatically infer text typed on a virtual keyboard \cite{ispy}. They use inexpensive cameras (such as those in smartphones), utilize the popup of keys when pressed and adopt computer vision techniques processing the recorded video in order to infer the corresponding key although the text in the video is illegible.

Xu {\em et al}. extend the work in \cite{ispy} and track the finger movement to infer input text \cite{XHW+::SeeingDouble::2013}. Their approach has five stages: in Stage 1, they use a tracking framework based on AdaBoost \cite{GGB::AdaBoost::2006} to track the location of the victim device in an image. In Stage 2, they detect the device's lines, use Hough transform to determine the device's orientation and align a virtual keyboard to the image. In Stage 3, they use Gaussian modeling to identify the ``fingertip" (not touched points as in our work) by training the pixel intensity. In Stage 4, RANSAC is used to track the fingertip trajectory, which is a set of line segments. If a line segment is nearly perpendicular to the touch screen surface, it implicates the stopping position and the tapped key. In Stage 5, they apply a language model to identify the text given the results from previous stages that produce multiple candidates of a tapped key. They use two cameras: Canon VIXIA HG21 camcorder with 12x optical zoom and Canon 60D DSLR with 400mm lens.

%


\section{Homography Based Attack against Touching Screen }
\label{problemDef}

In this section, we first introduce the basic idea of the computer vision based attack disclosing touch input via planar homography and then discuss each step in detail.
Without loss of generality, we use the four-digit passcode input on iPad as the example.

\subsection{Basic idea of attack}

Planar homography is a 2D projective transformation that relates two images of the same planar surface. Assume $p = (s, t, 1)$ is any point in an image of a 3D planar surface and $q = (s', t', 1)$ is the corresponding point in another image of the same 3D planar surface. The two images may be taken by the same camera or different cameras. There exists an invertible $3 \times 3$ matrix $\mathbf{H}$ (denoted as homography matrix),
\begin{eqnarray}
\label{eqn::H}
q = \mathbf{H} p.
\end{eqnarray}

\begin{figure}[!htp]
\centering
\includegraphics[width=3.6in]{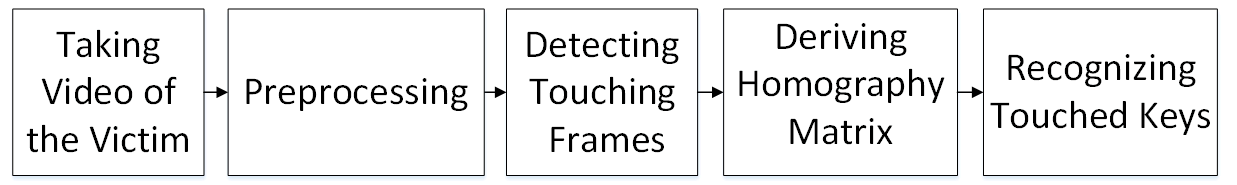}\\
\caption{Work flow of Homography-based Attack}
\label{overviewflow}
\end{figure}

\begin{figure}[!htp]
\begin{minipage}{1.3in}
\centering
\includegraphics[height=1in]{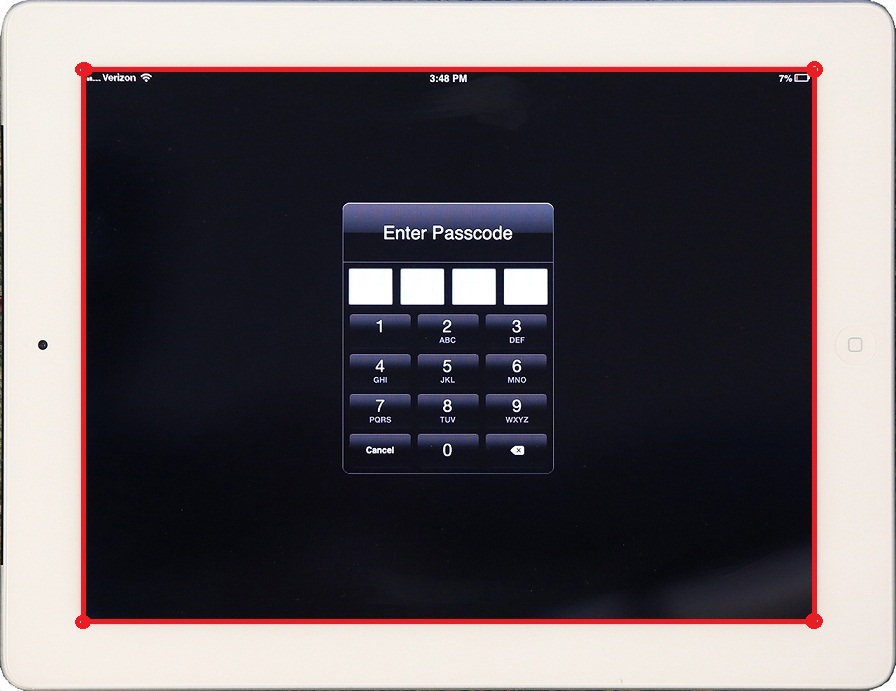}\\
\caption{Reference Image}
\label{destinationImage}
\end{minipage}
\begin{minipage}{2.1in}
\centering
\includegraphics[height=1in]{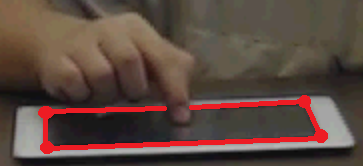}
\caption{Touching Frame} \label{touchingImage}
\end{minipage}
\end{figure}

Figure \ref{overviewflow} shows the basic idea of the attack.
{\bf Step 1.} From a distance, the attacker takes a video of the victim tapping on a device. We do not assume the video can show any text or popups on the device while we assume the video records finger movement on the touch screen surface.
{\bf Step 2.} We preprocess the video to keep only the area of touch screen with moving fingers. The type of device is known and we obtain a high resolution image of the virtual keyboard on this type of device, denoted as {\em reference image}, as shown in Figure \ref{destinationImage}.
{\bf Step 3.} Next, we detect video frames in which the finger touches the touch screen surface, denoted as {\em touching frames}, as shown in Figure \ref{touchingImage}. The touched position implicates the touched key.
{\bf Step 4.} Now we identify features of the touch screen surface, and derive the homography matrix between video frames and reference image. For example, we derive the homography matrix using Figures \ref{touchingImage} and \ref{destinationImage}.
{\bf Step 5.} Finally, we identify the touched points in the touching image and map them to the reference image via homography relationship in Equation (\ref{eqn::H}).
If the touched points can be correctly derived, we can disclose the corresponding touched keys. We introduce the five steps in detail below.

\subsection{Taking Video}
The attacker takes a video of a victim tapping on a divice from a distance. There are various such scenarios as students taking class, people attending conferences, and tourists gathering and resting in a square. Taking a video at such a place with a lot of people around should be stealthy. With the development of smartphones and webcams, such a stealthy attack is feasible. For example, iPhone has decent resolution. Galaxy S4 Zoom has a rear camera with 10x zoom and 16-megapixel, weighting only 208g. Amazon sells a webcam-like plugable USB 2.0 digital microscope with 2MP and 10x-50x optical zoom \cite{Plugable::Microscope::2013}.

In addition to the quality of the camera, three other factors affect the quality of the video and the result of recognized touched keys: angle, distance, and lighting. The basic idea of the attack is to identify touched points by the finger on the touch screen surface. The camera needs to take an angle to see the finger movement and the touch screen. For example, in a conference room, people in the front can use the front camera of their phone to record a person tapping in the back row. The distance also plays a critical role. If the camera is too far away from the victim, the area of the touch screen will be too small and the finger's movement on the screen will be hard to recognize. Of course, a camera with large zoom can help in case that the target is far. Lighting also plays an important role for recognizing the finger and touched points. It may affect the brightness and contrast of the video.

\subsection{Preprocessing}
\label{sectionPreprocessing}

In the step of preprocessing, we crop the video and keep only the area of touch screen with the moving hand. This removes most of the useless background since we are only interested in the touch screen surface where the finger touches keys. If the device does not move in the touching process, we just need to locate the area of the tablet in the first video frame and crop the same area for all the frames of the video.
If the device moves when the user inputs, we need to track its movement and crop the corresponding area. There are a lot of tracking methods \cite{objectTraSurvey}. We choose to use predator \cite{TLD}: we first draw the bounding box of the tablet area. The tracker will track its movement, and return its locations in every frame.

We are particularly interested in the fingertip area, where the finger touches the key. In general the resolution of this area is so poor that it is hard to identify. Therefore, we resize the cropped frames to add redundancy. For example, we resize each cropped frame to four times its original size.

We assume we know the type of device the victim uses and can get an image of the device with its touch screen area showing the virtual keyboard, denoted as ``reference image". It is easy to recognize most tablets and smartphones since each brand of device has salient features. For example, by passing the victim intentionally and glancing at the victim's device, we can easily get to know its type. We may also identify the brand from the video.
Once the device brand and model are known, we can get the reference image, whose quality is critical. The image shall show every feature of the device, particularly the planar surface of touch screen. For example, for iPad, we choose a black wallpaper so that the touch screen has a high contrast with its white frame. The virtual image of the camera shall not appear in the reference image in order to reduce noise in later steps.

\subsection{Detecting touching frames} \label{detectTouchFrame}
Touching frames are those video frames in which the finger touches the screen surface. To detect them, we need to analyze the finger movement pattern during the passcode input process. People usually use one finger to tap on the screen and input the passcode. We use this as the example to demonstrate the essence of our technique.

During the touching process, it is intuitive to see that the fingertip first moves downward towards the touch screen, stops, and then moves upward away from the touch screen. Of course the finger may also move left or right while moving downward or upward.
We define the direction of moving toward the device as positive and moving away from the device as negative.
Therefore, in the process of a key being touched, the fingertip velocity is first positive while moving downward, then zero while stopping on the pad and finally negative while moving forward. This process repeats for each touched key.
Therefore, a touching frame is the one where the velocity of the fingertip is zero. Sometimes the finger moves so fast that there is no frame where the fingertip has a velocity of zero. In such case, the touching frame is the one where the fingertip velocity changes from positive to negative.

The challenge to derive the velocity of the fingertip is to identify the fingertip in order to track its movement. The angle we take the video affects the shape of the fingertip in the video. Its shape changes when the soft fingertip touches the hard touch screen surface. People may also use different areas of the fingertip to tap the screen. Therefore, it is a challenge to automatically track the fingertip and identify the touching frames.

After careful analysis, we find that when people touch keys with the fingertip, the whole hand most likely keep the same gesture in the whole process and move in the same direction. Instead of tracking the fingertip movement to identify a touching frame, we can track the movement of the whole hand, or the whole finger touching the key. The whole hand or finger has enough feature for an automatic analysis.

We employ the theory of optical flow \cite{visionBOOK} to get the velocity of points on the moving finger or hand. Optical flow is a technique to compute object motion between two frames. The displacement vector of the points between the subsequent frames is called the image velocity or the optical flow at that point. We employ the KLT algorithm \cite{KLT}, which can track sparse points. To make the KLT algorithm effective, we need to select good and unique points, which are often corners in the image. The Shi-Tomasi corner detector \cite{goodfeature} is applied to get the points. We would track several points in case some points are lost during the tracking. If the velocity of most points change from positive to negative, this frame will be chosen as the touching frame. Our experiments show that six points are robust to detect all the touching frames.

From the experiments, we find that most of the time, for each touch with the finger pad, there are more than one touching frames. This is because the finger pad is soft. When it touches the screen, the pressure will force it to deform and this takes time. People may also intentionally stop to make sure that a key is touched. During the interaction, some tracked points may also move upward because of this force. We use a simple algorithm to deal with all the noise: if the velocity of most of the tracked points in one frame moves from positive to negative, that frame is a touching frame. Otherwise, the last frame where the finger interacts with the screen will be chosen as the touching frame.

\subsection{Deriving the Homography Matrix} \label{comHomoMatrix}
In computer vision, automatically deriving the homography matrix $H$ of a planar object in two images is a well studied problem \cite{multiViewGeomerty}. It can be derived as follows. First, a feature detector such as SIFT (Scale-Invariant Feature Transform) \cite{siftFeature} or SURF (Speeded Up Robust Features) \cite{surfFEATURE} is used to detect feature points. Matching methods such as FLANN (Fast Library for Approximate Nearest Neighbors) \cite{flanMATCHER} can be used to match feature points in the two images.
The pairs of matched points are then used to derive the homography matrix via the algorithm of
RANSAC (RANdom SAmple Consensus) \cite{ransac}.
However, those common computer vision algorithms for deriving $\mathbf{H}$ are not effective in our scenario. Because of the perspective of taking videos and reflection of touch screen, there are few good feature points in the touch screen images. Intuitively touch screen corners are potential good features, but they are blurry in our context since the video is taken remotely. SIFT or SURF cannot correctly detect those corners.

We derive the homography matrix $\mathbf{H}$ in Equation (\ref{eqn::H}) as follows. $\mathbf{H}$ has 8 degrees of freedom (Despite 9 entries in it, the common scale factor is not relevant). Therefore, to derive the homography matrix, we just need 4 pairs of matching points of the same plane in the touching frame and reference image. Any three of them should not be collinear \cite{multiViewGeomerty}.
In our case, we try to get the corners of the touch screen as shown in Figure \ref{touchingImage} and Figure \ref{destinationImage}.
Because the corners in the image are blurry, to derive the coordinates of these corners, we detect the four edges of the touch screen and the intersections of these edges are the desired corners. We apply the Canny edge detector \cite{Canny} to detect the edges and use the Hough line detector \cite{houghline} to derive possible lines in the image. We choose the lines aligned to the edges. Now we calculate intersection points and derive the coordinates of the four corners of interest.
With these four pairs of matching points, we can derive the homopgraphy matrix with the DLT (Direct Linear Transform) algorithm \cite{multiViewGeomerty} by using OpenCV \cite{learningOpencv}.

If the device does not move during the touching process, this homography matrix can be used for all the video frames. Otherwise, we should calculate $H$ for every touching frame and the reference image.

\subsection{Recognizing Touched Keys}

With the homography matrix, we can further determine what keys are touched. If we can determine the touched points in a touching image in Figure \ref{touchingImage}, we can then map them to the points in the reference image in Figure \ref{destinationImage}. The corresponding points in the reference image are denoted as mapped points. Such mapped points should land in the corresponding key area of the virtual keyboard in the reference image. Therefore, we derive the touched keys. This is the basic idea of blindly recognizing the touched keys although those touched keys are not visible in the video.
The key challenge is to determine the touched points. We propose the clustering-based matching strategy to address this challenge and will introduce it in Section \ref{sec::KeyMatching}.


\section{Recognizing Touched Keys}
\label{sec::KeyMatching}

To recognize touched keys, we need to identify the area where the finger touches the touch screen surface. In this section,
we analyze how people use their finger to tap and input text, and the image formation process of the finger touching the touch screen. We then propose a clustering-based matching strategy to match touched points in the touching frames and keys in the reference image.

\subsection{Formation of Touching Image}
\begin{figure}[!htp]
\begin{minipage}[t]{1.6in}
\centering
\includegraphics[width=1.4in]{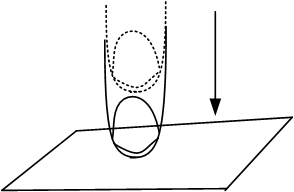}
\caption{Vertical Touching} \label{verticalTouching}
\end{minipage}
\begin{minipage}[t]{1.6in}
\centering
\includegraphics[width=1.6in]{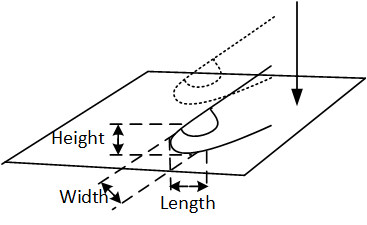}
\caption{Oblique Touching} \label{ObliqueTouching}
\end{minipage}
\end{figure}

To analyze how touching images are formed, we first need to analyze how people tap on the screen, denoted as {\em touching gestures}. According to \cite{gesture1, gesture2, gesture3}, there are two types of interaction between the finger and the touch screen: vertical touch and oblique touch. In the case of vertical touch, the finger moves downward vertically to the touch screen as shown in Figure \ref{verticalTouching}. People may also choose to touch the screen from an oblique angle as shown in Figure \ref{ObliqueTouching}, which is the most common touching gesture, particularly for people with long fingernails. The terms of vertical and oblique touch refer to the ``steepness" (also called ``pitch") difference of the finger \cite{gesture2}. From Figures \ref{verticalTouching} and \ref{ObliqueTouching}, the finger orientation (or `yaw') relative to the touch screen may also be different \cite{yaw}. Specific to every person and touch screen, the shape and size of a finger and key size also affect the touching gestures and where keys would be touched.

\begin{figure}[!htp]
\centering
\includegraphics[width=2.7in]{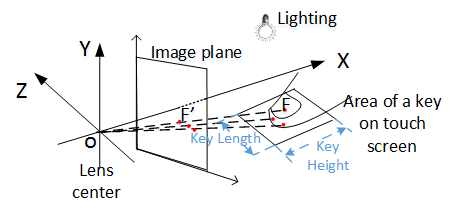}\\
\caption{Image Formation of Fingertip in 3D}
\label{figureTouc3D}
\end{figure}

Now we analyze how an image of a finger touching a key is formed. Videos may be taken from different angles. Without loss of generality, we study the case that the camera faces the touching finger. Figure \ref{figureTouc3D} shows the geometry of the image formation of the touching process in the 3D world when the fingertip falls inside the key area. The point $F$ on the fingertip will project to the point $F'$, which is the intersection of the ray $OF$ and the image plane. Its brightness in the image will be determined by illumination and the fingertip shape. Intuitively, because of the lighting difference, points on the side of the finger touching the surface are dark in the image. Adjacent to the dark area is the gray area where lighting is not sufficient. There is also the bright area on the finger that is well illuminated.

\begin{figure*}[th]
\begin{center}
\begin{minipage}{2.3in}
\centering
\includegraphics[height=1.8in]{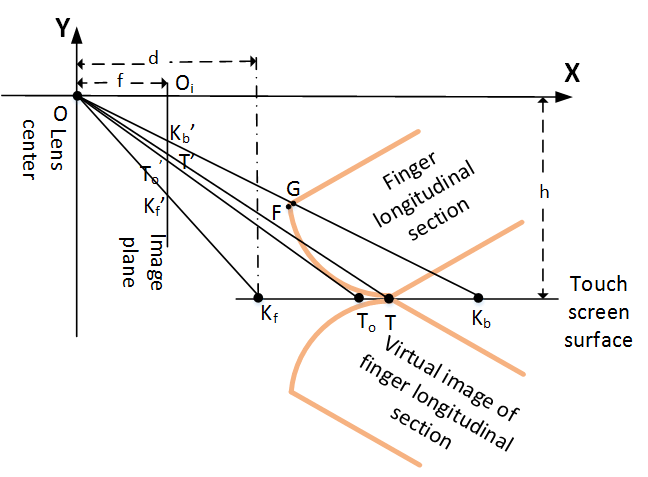}\\
\caption{Touching inside the Key}
\label{InkeyOblique}
\end{minipage}
\begin{minipage}{2.2in}
\centering
\includegraphics[height=1.8in]{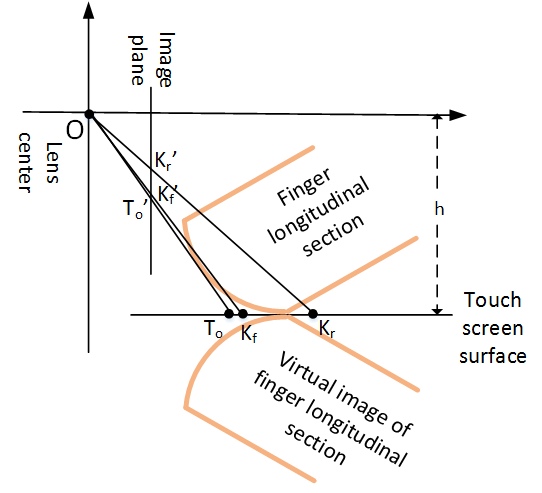}\\
\caption{Touching Point outside the Key}
\label{TouchingOutsideFu}
\end{minipage}
\begin{minipage}{2.2in}
\centering
\includegraphics[height=1.8in]{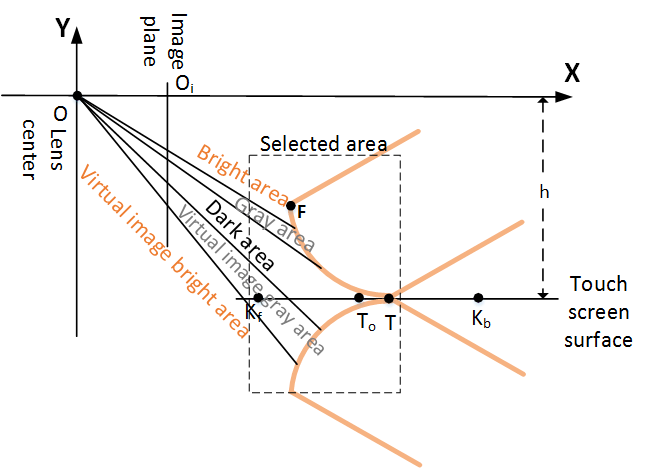}\\
\caption{Five Pixel Groups around Fingertip}
\label{fivePixelGroupsFu}
\end{minipage}
\end{center}
\end{figure*}

Figure \ref{InkeyOblique} shows the longitudinal view of a finger touching the surface and we will use it to discuss our basic principle to infer a touched key. $K_f$ and $K_r$ are the front and back of the touched key respectively. $T$ is the touched point. Apparently $T$ is on the line segment $\overline{K_fK_r}$. $T$ and $\overline{K_fK_r}$ are projected onto the image plane as $T'$ and $\overline{K_f'K_r'}$. If we can identify $T'$ in the image, our problem is solved. However, as we can see from Figure \ref{InkeyOblique}, since the human finger has a curved surface, the camera may not be able to see the touched point. $\overline{OT_o}$ is the tangent line to the curved finger surface and it intersects with the touch screen surface at $T_o$. The camera can see $T_o$, which is the closest point to the touched point on the touch screen surface. $T_o$ is projected as $T_o'$ on the image plane. If $T_o'$ is on the line segment $\overline{K_f'K_r'}$, then we just need to find $T_o'$ in the image and $T_o'$ can be used to determine the touched key.

We argue that $T_o'$ generally lands in the area of a key. Table \ref{tbl::KeySize} shows the key size of a virtual keyboard for iPad, iPhone and Nexus 7 tablet. Figure \ref{figureTouc3D} gives the definition of key height and length. Table \ref{tbl::FingertipSize} gives the average size of fingertip for index and middle fingers of 14 students of around 27 years old, including 4 females and 10 males. The fingertip height is the distance from the fingertip pulp to fingernail. The fingertip length is the distance between the fingertip pulp to the far front of the finger. When people touch the touch screen, they generally use the upper half of the fingertip to touch the middle of the key so that the key can be effectively pressed. We can see that half of the fingertip is around 6.5mm, less than the key height for all devices in Table \ref{tbl::KeySize}. Moreover, according to Tables \ref{tbl::KeySize} and \ref{tbl::FingertipSize}, the fingertip width is smaller than the key length. Therefore, the fingertip generally lands inside the key area, as shown in Figure \ref{InkeyOblique}. That is, the far front of the fingertip $F$ in Figure \ref{InkeyOblique} is in the range of the key and the touched point is inside the key area. Based on the perspective projection, $T_o$ is on the segment of $\overline{K_fK_b}$ so that $T_o'$ is on the segment of $\overline{K_f'K_b'}$ whenever the fingertip is in the view of the camera.

\begin{table}[h]
\caption{Virtual Keyboard Key Size}
\label{tbl::KeySize}
\centering
\begin{tabular}{|c|c|c|c|}
  \hline
   & iPad & iPhone 5 & Nexus 7 \\
  \hline
  Height (mm) $\times$ Length (mm) & $9 \times 17 $ & $8 \times 16$ & $10 \times 16$ \\
  \hline
\end{tabular}
\end{table}

\begin{table}[h]
\caption{Fingertip Size}
\label{tbl::FingertipSize}
\begin{tabular}{|c|c|c|c|c|}
\hline
 &\multicolumn{2}{l|}{Index Finger}&\multicolumn{2}{l|}{Middle Finger}\\
\cline{2-5}
 & Average & Standard Deviation & Average & Standard Deviation\\
\hline
Height (mm) & 9.6 & 1.2 & 10.4 & 1.3 \\
\hline
Length (mm) & 12.9 & 1.6 & 13.1 & 1.7 \\
\hline
Width (mm) & 13.1 & 1.9 & 13.7 & 1.7 \\
\hline
\end{tabular}
\end{table}

There are cases that $T_o'$ is not on the line segment $\overline{K_f'K_r'}$. Figure \ref{TouchingOutsideFu} illustrates such a case. Please note we intentionally draw a large finger for clarity. In this case, the key is so small. The camera is too close to the finger and takes such a wrong angle that $T_o$ lands outside $\overline{K_fK_r}$. Therefore, $T_o'$ is not on the line segment $\overline{K_f'K_r'}$. Even if we find $T_o'$, it gives the wrong key. However, we argue that keys of a touch screen virtual keyboard are designed to be large enough in order to hold the front of the fingertip as shown in Table \ref{tbl::KeySize}. The attacker also has the flexibility to choose the right angle so that the fingertip front is in the view of the camera. Another observation is that in the extreme case shown in Figure \ref{TouchingOutsideFu}, the touched key is behind the key holding $T_o$. Therefore, we may derive two candidate keys in the extreme case. In our experiments, we perform a second guess if the first guess is wrong because our technique can limit the touched keys to a few candidates. Our second time success rate is often very high.

We now derive the size of a key on an image and investigate its impact. The camera focus length is $f$. The height from the camera to the touch screen surface is $h$. The physical key size $|\overline{K_fK_b}|=w$. The distance between the key front $K_f$ and the lens center is $d$. By simple geometry operation, we have
\begin{eqnarray}
\label{eqn::keysize}
|K_f^{'} K_b^{'}| = \frac{fh}{d(1+d/w)}.
\end{eqnarray}
From (\ref{eqn::keysize}), the farther the touch screen from the camera, the smaller the size of the key in the image. The smaller the physical key size, the smaller of the key in an image. Table \ref{tbl::CameraSpecification} gives the camera specification of two cameras used in our experiments: Logitech HD Pro Webcam C920 \cite{camera} and the iPhone 5 camera. If the camera is around 2 meters away and half a meter away from the target, according to Equation (\ref{eqn::keysize}) and our experiments, the key size is only a few pixels. Therefore, in our experiments, we often need to zoom the fingertip for accurate localization of touched points.

\begin{table}[h]
\caption{Camera Specification}
\label{tbl::CameraSpecification}
\centering
\begin{tabular}{|c|c|c|}
  \hline
  Camera & Focal Length (mm) & Pixel Size ($\mu$m) \\
  \hline
  Logitech C920 & 3.67 & 3.98 \\
  \hline
  iPhone 5 & 4.10 & 1.40 \\
  \hline
\end{tabular}
\end{table}

\subsection{Clustering-based Matching}
\label{sectionCluster}
Based on the principle of blindly recognizing touched keys introduced above, we now introduce the clustering-based matching strategy recognizing touched keys.
If we can derive the position of the touched point $T_O^{'}$ in Figure \ref{fivePixelGroupsFu}, we can infer the corresponding key by applying homography. The problem is how to identify this touched point\footnote{Touching points actually form an area under the fingertip.}. Intuitively, since $T_O^{'}$ is far below the fingertip, which blocks lighting, $T_O^{'}$ should be in the darkest area around the fingertip in the image. Our experiments have verified this claim. However, it is hard to identify dark or gray.

We now analyze the brightness of the area around the fingertip in order to identify touched points.
The fingertip is a very rough surface at the microscopic level and can be treated as an ideal diffuse reflector. The incoming ray of light is reflected equally in all directions by the fingertip skin. The reflection conforms to the Lambert's Cosine Law \cite{visionBOOK}: the reflected energy from a small surface area in a particular direction is proportional to cosine of the angle between the particular direction and the surface normal.
Therefore, for the lower part of the fingertip arc facing the touch screen, denoted as inner side of the fingertip, the angle is large and less energy will be reflected so that the pixels are darker.
Particularly, the area around $T_O^{'}$ is the darkest. The area around the fingertip top $F$ is the brightest. From the bright area to dark area, there exists the gray area between $F$ and $T_O^{'}$. Since the touch screen is basically a mirror, the camera may also capture the virtual image of the inner side of the fingertip, which also has a bright area, gray area and bright area.

Therefore, around the fingertip and its virtual image, we can have five areas with five different brightness: bright fingertip top, gray fingertip middle area, dark fingertip bottom, dark fingertip bottom of the virtual image, gray fingertip middle area of the virtual image and bright fingertip top of the virtual image. $T_O^{'}$ lands in the upper half part of the dark area since the other half of the dark area is formed by the virtual image of dark fingertip bottom.

\begin{figure}[!htp]
\begin{center}
\begin{minipage}[t]{1.7in}
\centering
\includegraphics[height=1.5in]{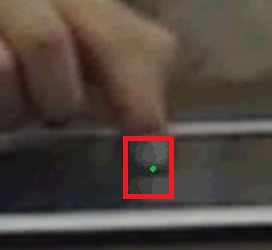}
\caption{Clustered Result and the Chosen Point} \label{clusterred}
\end{minipage}
\begin{minipage}[t]{1.7in}
\centering
\includegraphics[height=1.5in]{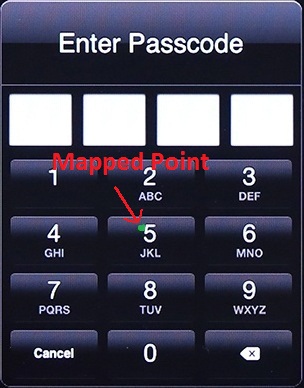}
\caption{Mapped Point} \label{clusterResult}
\end{minipage}
\end{center}
\end{figure}

We can use clustering algorithms to cluster these five areas of pixels with different brightness. We select a small rectangle area around the fingertip and its virtual image and apply the k-means clustering to pixels in this area. The number of clusters is set as $5$. Then the darkest cluster refers to the area where the finger touches the screen surface. We select pixels in the upper half of this cluster and apply homography to find the corresponding keys. Sometimes, the lighting and shading may disturb our claim that the darkest cluster is where the finger touches the screen. However, because the area around the touched point has quite different intensity from other areas, we can still use the cluster around the fingertip top to infer the touched keys. Basically, the clustering algorithm helps {\em accurately} identify the touched area. As an example, the (red) box in Figure \ref{clusterred} shows the clustered result of the selected area, and the selected point (green dot). Figure \ref{clusterResult} shows the mapped point (in green) that falls into the area of Key five. Therefore, five is the touched key.

If we examine Figure \ref{InkeyOblique} carefully, we can see that in addition to touched points, points on the fingertip arc may also be projected into the key area $\overline{K_f'F_b'}$. In Figure \ref{InkeyOblique}, line $\overline{OK_b}$ and the finger intersect at point $G$. We can see that all points on the fingertip arc visible to the camera are projected into the area of the key in the image. Therefore, in this case, both touched points and points on the fingertip arc can be used to deduce the key even if the points on fingertip arc are in the bright or gray area from our clustering above. However, due to the size of the finger, touched position, angle of touching, distance, height and angle of the camera, the position of $G$ changes too and can be any point on the fingertip arc. It is not reliable to use these points on the fingertip arc to infer the touched key. We still use touched points in the dark area, but the fact that points in the gray or bright area may be projected into the key's area lends us some robustness to use touched points in the dark area to infer the touched key.

\section{Evaluation} \label{evaluation}

In this section, we present the experimental design and results.
To demonstrate the impact of the attack, we have performed extensive experiments on various devices with different key size, including iPad, iPhone and Nexus 7 tablet. Two cameras are used: Logitech HD Pro Webcam C920 \cite{camera} and the iPhone 5 camera, whose specification is in Table \ref{tbl::CameraSpecification}. All the experiments are done with the Logitech HD Pro Webcam C920 except for the last group of experiments for comparison between web camera and phone camera.

\subsection{Experimental Design}
\label{sec::expDesign}
We consider the following factors during the experimental design: users, camera distance from the target device, and angles of the camera. Passcodes in the experiments are randomly generated.

\begin{figure}[!htp]
\centering
\includegraphics[width=2.5in]{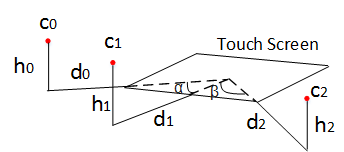}\\
\caption{Experimental Setting}
\label{experimentSetting}
\end{figure}

{\bf Users}:
Different people have different finger shape, fingernail and touching gestures. Five females and six males with the experience of using tablets and smartphones participated in the experiments. They are separated to three groups: 3 people in the first group, 7 people in the second group. These two groups perform experiments with iPad. The last group helps us to evaluate the success rate versus the distance between the camera and target.
For the first group, we take 10 videos for every person per angle (front, left front and right front of the target device) as shown in Figure \ref{experimentSetting}, and 90 videos are taken. For the second group, five videos are taken for every person per angle and 105 videos are taken.
Totally, 195 videos are taken. During the experiments, users tap in their own way without any restriction.

{\bf Angles and Distance}:
Even for the same person, the touching gestures appear to be different in videos taken from different distances and angles. As shown in Figure \ref{experimentSetting}, we take videos from three angles (front, left front and right front) and at different distances. In Figure \ref{experimentSetting}, $C_0$, $C_1$, and $C_2$ are the cameras, $d_0$, $d_1$, and $d_2$ are the lateral distance, $h_0$, $h_1$, and $h_2$ are the vertical distance from the camera to the device, and $\alpha$ and $\beta$ are the angles between the camera and the device. The angle can be adjusted to make the touching fingertip show up in the video.
For the first two groups, videos are taken from 2.1m to 2.4m away, height about 0.5m, and $\alpha$ and $\beta$ around 40 degree. To test how the distance affects the results, we take some videos in the front of the target with distance of 2m, 3m, 4m and 5m, and height about 1m.

{\bf Lighting}:
The lighting will affect the brightness and contrast of the image. The experiments are performed in a classroom with dimmable lamps on the ceiling. Videos in the first group are taken under a normal light, and experiments of the second group are taken under a much stronger light. Other experiments are performed under normal light.


\subsection{Detecting Touching Frames via Optical Flow}

As discussed in Section \ref{detectTouchFrame}, we track feature points and use their velocity change to detect touching frames. To confirm how many feature points should be used, we apply different number of feature points to videos.
Our experiments show that 5 or more feature points are stable for tracking touching frames with a true positive of 100\% as shown in Table \ref{detectTouchingRatio}. The optical flow algorithm may also produce false positive, falsely recognizing frames in which a finger does not touch the screen surface as touching frames. The false positive rate is very low, less than 1\%, as shown in Table \ref{detectTouchingRatio}. Assume that the passcode length is 4. If more than 4 frames are detected as touching frames, we just manually remove the extra frames. It is often obvious that the extra frames are not touching frames.

\begin{table}[h]
\caption{Detecting Touching Frame Result}
\centering
\begin{tabular}{|c|c|c|c|c|}
  \hline
   \backslashbox{Rate}{Perspective} & Front & Left Front & Right Front & Total \\
   \hline
  True Positive &  100\% &  100\%  &  100\% &  100\% \\
  \hline
  False Positive &  0.91\%& 0.88\% & 0.88\% & 0.89\% \\
  \hline
\end{tabular}
\label{detectTouchingRatio}
\end{table}

\subsection{Recognizing Touched Keys on iPad via Webcam}
Table \ref{oneTimekeyResult} shows the success rate of the retrieved keys from videos taken from different angles. We define success rate as the ratio of the number of the correctly retrieved passcodes (all four digits) over the number of videos. For the wrong results, we give a second try, choosing other candidates of passcode digits. It can be observed that the overall first-time success rate from different angles reaches more than 90\%.  The second time success rate is higher than the first time success rate and reaches over 95\%.

\begin{table}[h]
\caption{Success Rate of Clustering Based Matching}
\centering
\begin{tabular}{|c|c|c|c|c|}
  \hline
   \backslashbox{Success Rate}{Angle} & Front & Left Front & Right Front & Total \\
   \hline
  First Time & 90.8\% &87.9\%  & 95.2\% & 91.2\% \\
  \hline
  Second Time & 98.5\% &90.9\%  & 96.8\% & 95.3\% \\
  \hline
\end{tabular}
\label{oneTimekeyResult}
\end{table}

There are one or two wrong keys in most of the failed experiments. As analyzed in Section \ref{sec::KeyMatching}, for each touch we can produce two candidates. Thus, we would possibly correct the wrong keys in the second time try. This is why the second time success rate is higher than the first time success rate.

\begin{figure}[!htp]
\centering
\includegraphics[width=2.5in]{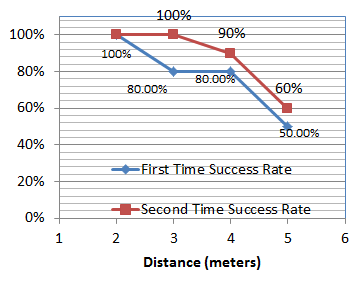}
\caption{Success Rate v.s. Distance} \label{fig::SuccessRateVSDistance}
\end{figure}

To check how the distance affects success rate, we take 10 videos for distance 2m, 3m, 4m and 5m respectively, with the Logitech HD Pro Webcam C920 spying iPad. From the results shown in Figure \ref{fig::SuccessRateVSDistance}, it can be observed that as distance increases, success rate decreases. At 5m, the first time success rate is around 50\%. This is because at such a distance the key size in the image is very small and can be up to 1 pixel wide. It is much more difficult to distinguish the touched key at the long distance. However, a camera with high optical zoom shall help. That is why such cameras are used in related work to spy from as far as 30 meters. However, our threat model does not allow the use of those high zoom cameras.

\subsection{Comparing Different Targets and Cameras}
\begin{figure}[!htp]
\centering
\includegraphics[height=1.8in]{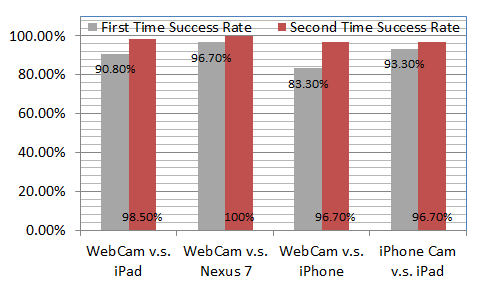}
\caption{Success Rate Comparison} \label{fig::SuccessRateComparison}
\end{figure}

To compare the success rate of recognizing touched keys on different devices, we perform thirty experiments for Nexus 7 and iPhone 5 respectively with Logitech HD Pro Webcam C920, two meters away and about about 0.65m high in the front of the device. To compare the success rate achieved by different cameras, we conducted thirty experiments with iPhone 5 recording tapping on iPad, from similar distance and height. Figure \ref{fig::SuccessRateComparison} shows the results by the clustering based attack. We also include the result of Logitech C920 versus iPad for comparison.
We can see that recognition of touched keys on iPhone 5 has a first time success rate of 83.30\%, below 90\%. The reason is due to the iPhone 5's keys being smaller. It is also harder to detect edges of the white iPhone 5 and the derived homography matrix may not be as accurate as those in other cases. The success rate is more than 90\% in all other cases. In all cases, the second time success rate is more than 95\%. The high success rate for all the cases demonstrates the severity of our attack.

%
%

\section{Countermeasures} \label{countermeasures}

In this section, we discuss countermeasures to computer vision based attacks introduced in this paper and related work. There are many other authentication approaches immune to these attacks to some extent, including biometric-rich gesture-based authentication \cite{Sae-Bae::Gestures::CHI2012, Yan::Leakage-Resilient::AsiaCCS2013, Kim::authentication::CHI2010} and graphic password schemes \cite{Biddle::GraphicalPasswords::Surveys2012, Suo::GraphicalPasswords::ACSAC2005, Bulling::Gaze-Based::CHI2012}.
The idea of randomized keyboard has been proposed for legacy keypad and touch-enabled devices \cite{Hirsch::Securekeyboard::1982, Hirsch::Securekeyboard::1984, McIntyre+::securepin::2003, Hoanca::Screentechnique::2005, Shin::RandomKeypad::2010, Lee::RandomKeypad::2011, Kim::RandomKeypad::2012}.
We have designed and developed context aware Privacy Enhancing Keyboards (PEK) for Android systems for the first time. PEK is a type of randomized keyboard. We have implemented PEK as a third party app and are also able to change the internal system keyboard to PEK.

\subsection{Design and Implementation of PEK}
A software keyboard may contain three sub-keyboards. The primary sub-keyboard is the QWERTY keyboard, which is the most common keyboard layout. The second sub-keyboard is the numerical keyboard that may also contain some symbols. The last sub-keyboard is a symbol keyboard that contains special symbols. The layout for these three sub-keyboards is stored in a XML file, which records the positions of keys and corresponding key codes. The system generates its keyboard in this way: the keys will be read from the XML file and put in a right position.

\begin{figure}
\begin{minipage}{1.7in}
  \centering
  \includegraphics[height=1.6in]{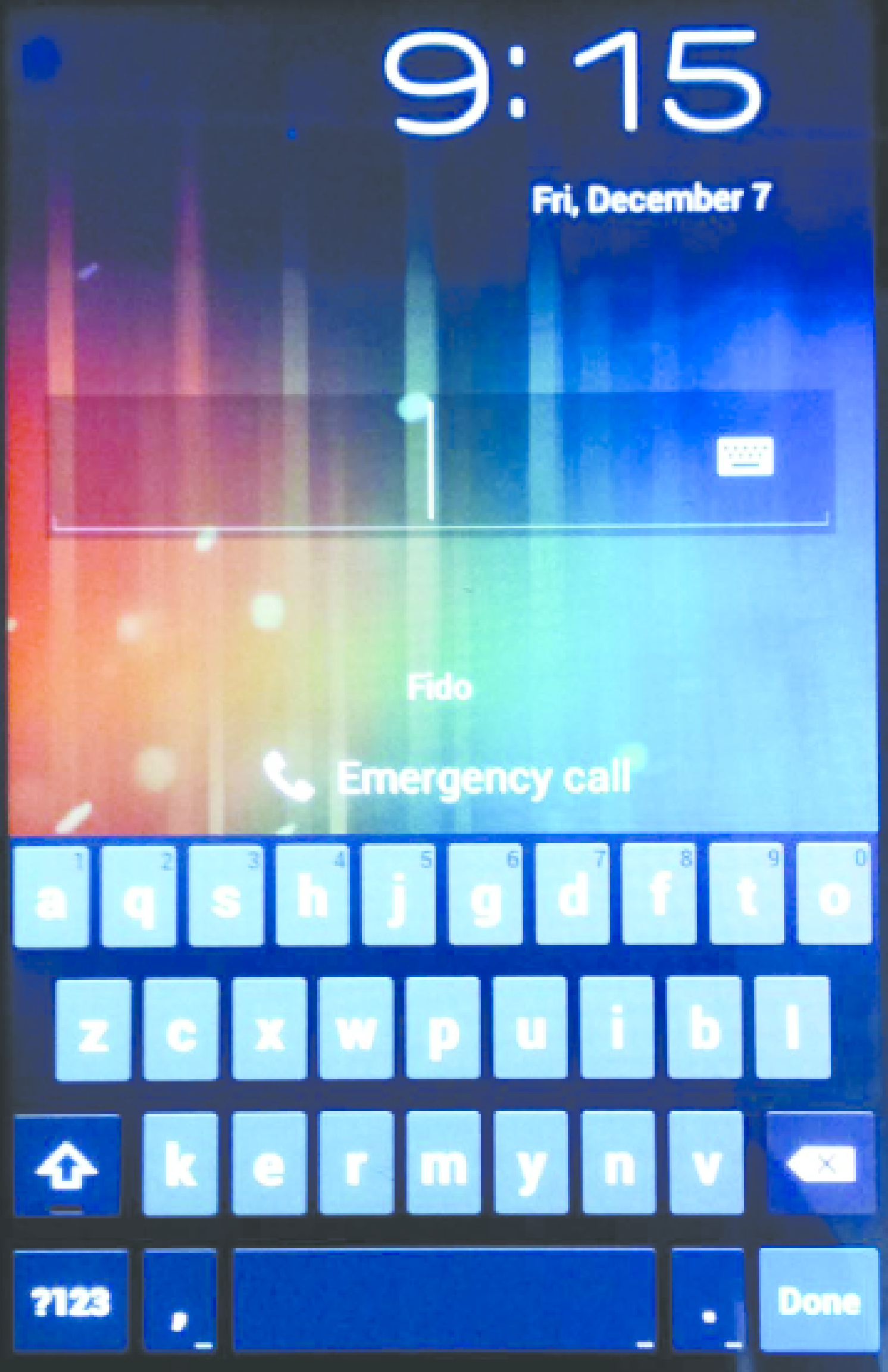}\\
  \caption{PEK: Shuffled Keys}\label{fig::Randomkeyboard}
\end{minipage}
\begin{minipage}{1.7in}
  \centering
  \includegraphics[height=1.6in]{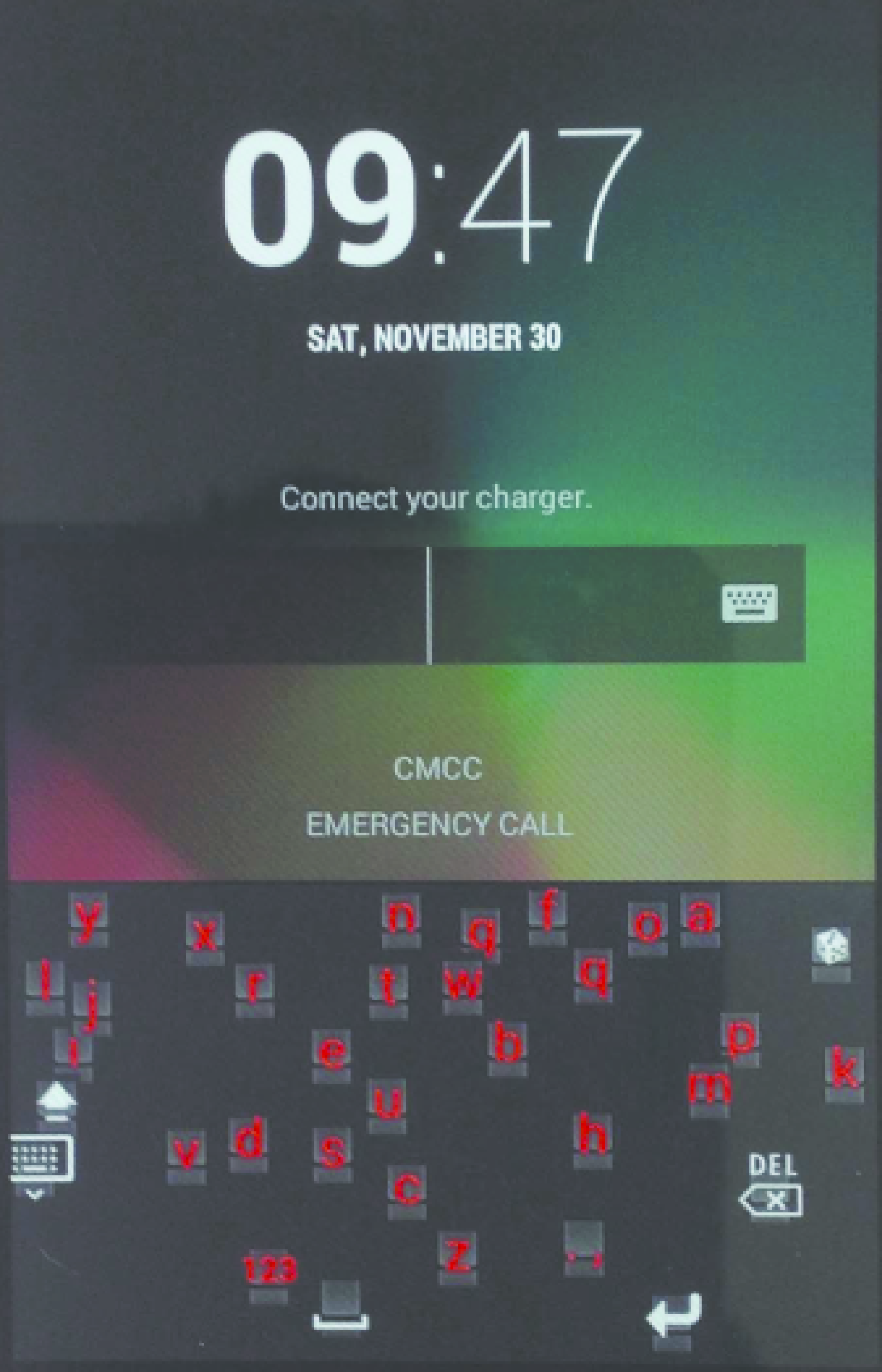}\\
  \caption{PEK: Brownian Motion}\label{fig::SizeRandomKeyboard}
\end{minipage}
\end{figure}

PEK changes the key layout to implement randomized keys. When a keyboard is needed, we first generate a random sequence of key labels for each of the three different keyboards. When a key is read from the XML, we randomly choose an integer number between one and the size of the key sequence. We use this number to pick the specific key label from the randomized key sequence and also remove it from the key sequence. This randomly selected key replaces the current key. In this way, we can manage to shuffle the key positions on a normal keyboard.
Another version of PEK is to make each key move within the keyboard region in a Brownian motion fashion by updating each key's position repeatedly according to Brownian motion. In this way, the keys are moving all the time. Even if the same key is pressed a few times, their positions are different. This is an improvement compared with PEK with shuffled keys in which the keyboard does not change in one session of password input. Figure \ref{fig::Randomkeyboard} show PEK with shuffled keys and Figure \ref{fig::SizeRandomKeyboard} shows PEK with Brownian motion of keys.

PEK is aware of the context and can pop up the randomized keyboard only if the input box is for sensitive information. The Android class ``EditorInfo'' can be used to detect the type of input box. In our case, ``TYPE\_NUMBER\_VARIATION\_PASSWORD'', ``TYPE\_TEXT\_VARIATION\_PASSWORD'', ``TYPE\_TEXT\_VARIATION\_VISIBLE\_PASSWORD'' and ``TYPE\_TEXT\_VARIATION\_WEB\_PASSWORD'' is used to identify the password input box. The first type is the variations of ``TYPE\_CLASS\_NUMBER'', while the last three types are the variations of ``TYPE\_CLASS\_TEXT''. Once the password input box is triggered by the user, a new randomized keyboard will be constructed. As a result, the user can have different key layouts every time she presses the password input box.

\subsection{Evaluation of PEK}
\begin{figure}[!htp]
\centering
\includegraphics[width=2.5in]{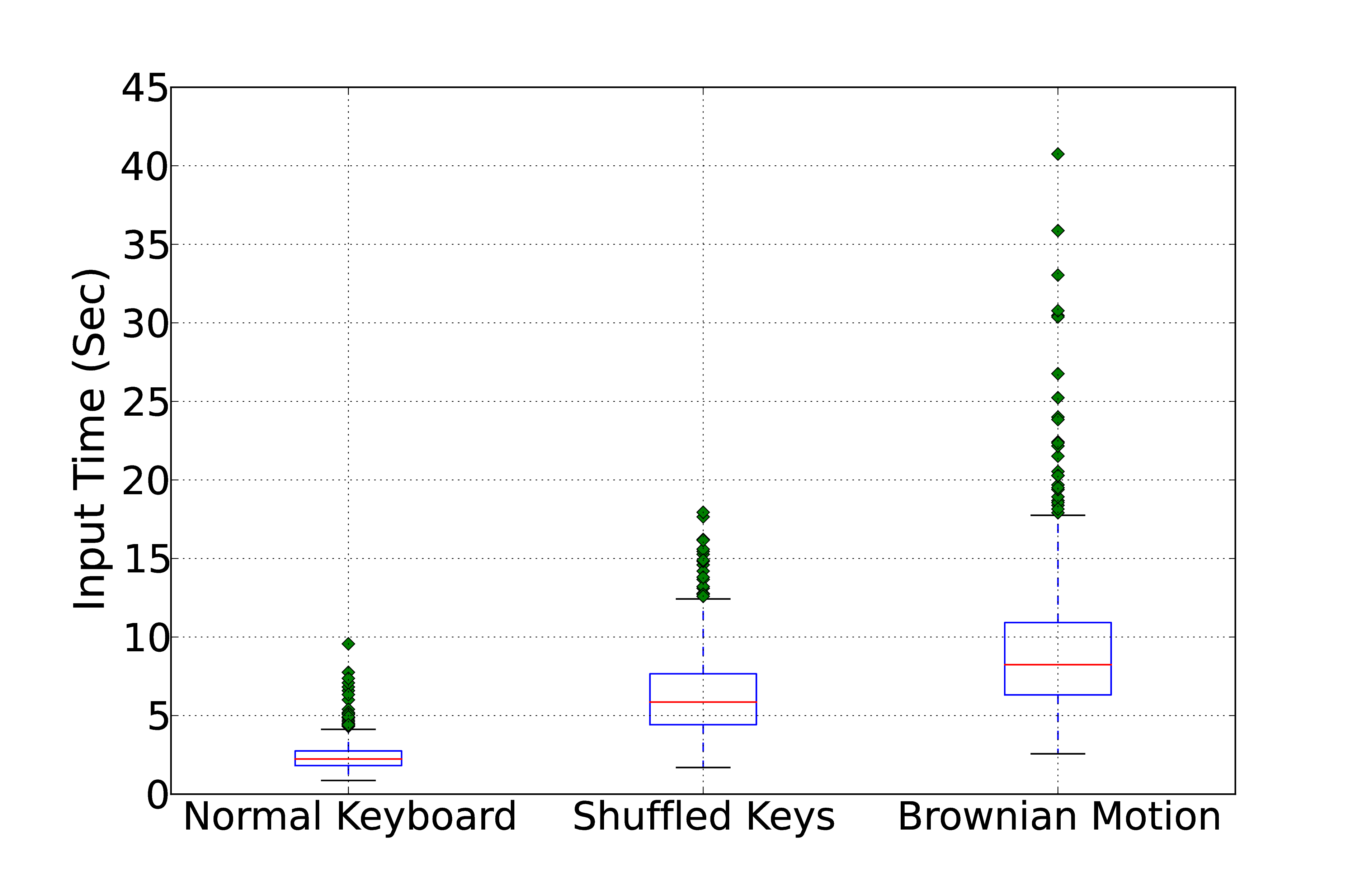}\\
\caption{Input Time of Three Distinct Keyboards}
\label{fig::usability}
\end{figure}

To measure the usability of our PEK, we recruit 20 students, 5 female students and 15 male students, whose average age is 25 years old. We implemented a test password input box and generated 30 random four-letter passwords. The students are required to input these 30 passwords by using a shuffled keyboard and a Brownian motion keyboard, and the test app records the user input time. Table \ref{usability} shows the results of our evaluation and Figure \ref{fig::usability} gives a box plot of input time of three different keyboards. The median input time is around 2.2 seconds on normal keyboard, 5.9 seconds on shuffled keyboard and 8.2 seconds on Brownian motion. Success rate is the probability that a user correctly inputs four-letter password. Success rate of all three keyboards are high while it is a little bit lower for the Brownian motion keyboard. The participants in our experiment feel PEK is acceptable if PEK only pops up the randomized keyboard for sensitive information input.

\begin{table}[th]
\caption{Usability Test}
\centering
\begin{tabular}{l|c|c|c}
  \hline
  & Normal & Shuffled  & Brownian\\
  & Keyboard & Keys & Motion\\
  \hline  Median Input Time (Second) & 2.235 & 5.859 & 8.24  \\
  \hline  Success Rate & 98.50\%  & 98.83\%  & 94.17\%\\
  \hline
\end{tabular}
\label{usability}
\end{table}

\section{conclusion} \label{conclusion}

In this paper, we present an attack that blindly recognizes input on touch screen from a distance.
The attack exploits the homography relationship between touching images where fingers touch the screen surface and reference image showing a virtual keyboard.
We use the optical flow algorithm to detect touching frames and designed the clustering matching strategy to recognize the touched keys. The homography matrix is derived by utilizing the intersections of the edges of the touch screen display. Our extensive experiments show that the first time success rate of recognizing touched keys is more than 90\% while the second time success rate is more than 95\%.
To defeat this type of attack, we design a context aware Privacy Enhancing Keyboard (PEK) which pops up a randomized keyboard on Android systems for sensitive information input such as passwords. PEK may use shuffled key or display a fully dynamic keyboard with keys moving in a Brownian motion pattern. Our experiments show that although PEK increases the passcode input time, it is acceptable for the sake of security and privacy to those we interviewed.

%

\bibliographystyle{IEEEtran}
\bibliography{visionattack}

\begin{thebibliography}{10}
\providecommand{\url}[1]{#1}
\csname url@samestyle\endcsname
\providecommand{\newblock}{\relax}
\providecommand{\bibinfo}[2]{#2}
\providecommand{\BIBentrySTDinterwordspacing}{\spaceskip=0pt\relax}
\providecommand{\BIBentryALTinterwordstretchfactor}{4}
\providecommand{\BIBentryALTinterwordspacing}{\spaceskip=\fontdimen2\font plus
\BIBentryALTinterwordstretchfactor\fontdimen3\font minus
  \fontdimen4\font\relax}
\providecommand{\BIBforeignlanguage}[2]{{%
\expandafter\ifx\csname l@#1\endcsname\relax
\typeout{** WARNING: IEEEtran.bst: No hyphenation pattern has been}%
\typeout{** loaded for the language `#1'. Using the pattern for}%
\typeout{** the default language instead.}%
\else
\language=\csname l@#1\endcsname
\fi
#2}}
\providecommand{\BIBdecl}{\relax}
\BIBdecl

\bibitem{touchfuture}
D.~Lee, ``The state of touch-screen panel market in 2011,'' ser.
  \url{http://www.walkermobile.com/March_2011_ID_State_of_the_Touch_Screen_Market.pdf},
  2011.

\bibitem{JuniperMTR::2013}
``Juniper networks third annual mobile threats report,''
  \url{http://www.juniper.net/us/en/local/pdf/additional-resources/3rd-jnpr-mobile-threats-report-exec-summary.pdf},
  2013.

\bibitem{backes08compromising}
M.~Backes, M.~D\"{u}rmuth, and D.~Unruh, ``Compromising reflections or how to
  read lcd monitors around the corner,'' in \emph{Proceedings of IEEE Symposium
  on Security and Privacy}, 2008, pp. 158--169.

\bibitem{backes09tempest}
M.~Backes, T.~Chen, M.~Duermuth, H.~Lensch, and M.~Welk, ``Tempest in a teapot:
  Compromising reflections revisited,'' in \emph{Proceedings of 30th IEEE
  Symposium on Security and Privacy}, 2009, pp. 315--327.

\bibitem{clearShot}
D.~Balzarotti, M.~Cova, and G.~Vigna, ``Clearshot: Eavesdropping on keyboard
  input from video,'' in \emph{Proceedings of the 2008 IEEE Symposium on
  Security and Privacy}, ser. SP '08, 2008, pp. 170--183.

\bibitem{eavesSmartphone}
F.~Maggi, S.~Gasparini, and G.~Boracchi, ``A fast eavesdropping attack against
  touchscreens,'' in \emph{IAS}.\hskip 1em plus 0.5em minus 0.4em\relax IEEE,
  2011, pp. 320--325.

\bibitem{ispy}
R.~Raguram, A.~M. White, D.~Goswami, F.~Monrose, and J.-M. Frahm, ``ispy:
  automatic reconstruction of typed input from compromising reflections,'' in
  \emph{Proceedings of the 18th ACM conference on Computer and communications
  security}, ser. CCS '11, 2011, pp. 527--536.

\bibitem{XHW+::SeeingDouble::2013}
Y.~Xu, J.~Heinly, A.~M. White, F.~Monrose, and J.-M. Frahm, ``Seeing double:
  Reconstructing obscured typed input from repeated compromising reflections,''
  in \emph{Proceedings of the 20th ACM Conference on Computer and
  Communications Security {(CCS}}, 2013.

\bibitem{KLT}
J.~yves Bouguet, ``Pyramidal implementation of the lucas kanade feature
  tracker,'' \emph{Intel Corporation, Microprocessor Research Labs}, 2000.

\bibitem{siftFeature}
D.~G. Lowe, ``Distinctive image features from scale-invariant keypoints,''
  \emph{Int. J. Comput. Vision}, vol.~60, no.~2, pp. 91--110, Nov. 2004.

\bibitem{iphoneVibration}
P.~Marquardt, A.~Verma, H.~Carter, and P.~Traynor, ``(sp)iphone: decoding
  vibrations from nearby keyboards using mobile phone accelerometers,'' in
  \emph{Proceedings of the 18th ACM conference on Computer and communications
  security}, ser. CCS '11, 2011, pp. 551--562.

\bibitem{timingAttack}
D.~Foo~Kune and Y.~Kim, ``Timing attacks on pin input devices,'' in
  \emph{Proceedings of the 17th ACM conference on Computer and communications
  security}, ser. CCS '10, 2010, pp. 678--680.

\bibitem{smudgeSmartphone}
A.~J. Aviv, K.~Gibson, E.~Mossop, M.~Blaze, and J.~M. Smith, ``Smudge attacks
  on smartphone touch screens,'' in \emph{Proceedings of the 4th USENIX
  conference on Offensive technologies}, ser. WOOT'10, 2010, pp. 1--7.

\bibitem{fingerprintattack}
Y.~Zhang, P.~Xia, J.~Luo, Z.~Ling, B.~Liu, and X.~Fu, ``Fingerprint attack
  against touch-enabled devices,'' in \emph{Proceedings of the second ACM
  workshop on Security and privacy in smartphones and mobile devices}, ser.
  SPSM '12, 2012, pp. 57--68.

\bibitem{thermal}
M.~Zalewski, ``Cracking safes with thermal imaging,'' ser.
  \url{http://lcamtuf.coredump.cx/tsafe/}, 2005.

\bibitem{thermalAnalysis}
K.~Mowery, S.~Meiklejohn, and S.~Savage, ``Heat of the moment: characterizing
  the efficacy of thermal camera-based attacks,'' in \emph{Proceedings of the
  5th USENIX conference on Offensive technologies}, ser. WOOT'11, 2011, pp.
  6--6.

\bibitem{Hao::TouchLogger::HotSec2011}
L.~Cai and H.~Chen, ``{TouchLogger}: Inferring keystrokes on touch screen from
  smartphone motion,'' in \emph{Proceedings of the 6th USENIX Workshop on Hot
  Topics in Security (HotSec)}, 2011.

\bibitem{OHDPZ::ACCessory::2012}
E.~Owusu, J.~Han, S.~Das, A.~Perrig, and J.~Zhang, ``Accessory: Keystroke
  inference using accelerometers on smartphones,'' in \emph{Proceedings of The
  Thirteenth Workshop on Mobile Computing Systems and Applications
  (HotMobile)}.\hskip 1em plus 0.5em minus 0.4em\relax ACM, February 2012.

\bibitem{XBZ::TapLogger::2012}
Z.~Xu, K.~Bai, and S.~Zhu, ``Taplogger: Inferring user inputs on smartphone
  touchscreens using on-board motion sensors,'' in \emph{Proceedings of The ACM
  Conference on Wireless Network Security ({WiSec})}, 2012.

\bibitem{GGB::AdaBoost::2006}
H.~Grabner, M.~Grabner, and H.~Bischof, ``Real-time tracking via on-line
  boosting,'' in \emph{Proceedings of the British Machine Vision Conference},
  2006.

\bibitem{Plugable::Microscope::2013}
Plugable, ``Plugable usb 2.0 digital microscope for windows, mac, linux (2mp,
  10x-50x optical zoom, 200x digital magnification),''
  \url{http://www.amazon.com/Plugable-Digital-Microscope-Windows-Magnification/dp/B00AFH3IN4/ref=sr_1_1?ie=UTF8&qid=1382796731&sr=8-1&keywords=optical+zoom+webcam},
  2013.

\bibitem{objectTraSurvey}
A.~Yilmaz, O.~Javed, and M.~Shah, ``Object tracking: A survey,'' \emph{ACM
  Comput. Surv.}, vol.~38, no.~4, Dec. 2006.

\bibitem{TLD}
Z.~Kalal, K.~Mikolajczyk, and J.~Matas, ``Tracking-learning-detection,''
  \emph{IEEE Trans. Pattern Anal. Mach. Intell.}, vol.~34, no.~7, pp.
  1409--1422, Jul. 2012.

\bibitem{visionBOOK}
R.~Szeliski, \emph{Computer Vision: Algorithms and Applications}, 1st~ed.\hskip
  1em plus 0.5em minus 0.4em\relax Springer-Verlag New York, Inc., 2010.

\bibitem{goodfeature}
J.~Shi and C.~Tomasi, ``Good features to track,'' Tech. Rep., 1993.

\bibitem{multiViewGeomerty}
R.~Hartley and A.~Zisserman, \emph{Multiple View Geometry in Computer Vision},
  2nd~ed.\hskip 1em plus 0.5em minus 0.4em\relax Cambridge University Press,
  2003.

\bibitem{surfFEATURE}
H.~Bay, A.~Ess, T.~Tuytelaars, and L.~Van~Gool, ``Speeded-up robust features
  (surf),'' \emph{Comput. Vis. Image Underst.}, vol. 110, no.~3, pp. 346--359,
  Jun. 2008.

\bibitem{flanMATCHER}
M.~Muja and D.~G. Lowe, ``Fast approximate nearest neighbors with automatic
  algorithm configuration,'' in \emph{In VISAPP International Conference on
  Computer Vision Theory and Applications}, 2009, pp. 331--340.

\bibitem{ransac}
P.~Huber, \emph{Robust Statistics}.\hskip 1em plus 0.5em minus 0.4em\relax John
  Wiley \& Sons, 1981.

\bibitem{Canny}
J.~Canny, ``A computational approach to edge detection,'' \emph{IEEE Trans.
  Pattern Anal. Mach. Intell.}, vol.~8, no.~6, pp. 679--698, 1986.

\bibitem{houghline}
J.~Matas, C.~Galambos, and J.~Kittler, ``Robust detection of lines using the
  progressive probabilistic hough transform,'' \emph{Comput. Vis. Image
  Underst.}, vol.~78, no.~1, pp. 119--137, 2000.

\bibitem{learningOpencv}
G.~R. Bradski and A.~Kaehler, \emph{Learning opencv, 1st edition},
  1st~ed.\hskip 1em plus 0.5em minus 0.4em\relax O'Reilly Media, Inc., 2008.

\bibitem{gesture1}
H.~Benko, A.~D. Wilson, and P.~Baudisch, ``Precise selection techniques for
  multi-touch screens,'' in \emph{Proceedings of the SIGCHI Conference on Human
  Factors in Computing Systems}, ser. CHI '06, 2006, pp. 1263--1272.

\bibitem{gesture2}
C.~Forlines, D.~Wigdor, C.~Shen, and R.~Balakrishnan, ``Direct-touch vs. mouse
  input for tabletop displays,'' in \emph{Proceedings of the SIGCHI Conference
  on Human Factors in Computing Systems}, ser. CHI '07, 2007, pp. 647--656.

\bibitem{gesture3}
F.~Wang, X.~Cao, X.~Ren, and P.~Irani, ``Detecting and leveraging finger
  orientation for interaction with direct-touch surfaces,'' in
  \emph{Proceedings of the 22nd annual ACM symposium on User interface software
  and technology}, ser. UIST '09, 2009, pp. 23--32.

\bibitem{yaw}
F.~Wang and X.~Ren, ``Empirical evaluation for finger input properties in
  multi-touch interaction,'' in \emph{Proceedings of the SIGCHI Conference on
  Human Factors in Computing Systems}, ser. CHI '09, 2009, pp. 1063--1072.

\bibitem{camera}
Logitech, ``Logitech hd pro webcam c920,''
  \url{http://www.logitech.com/en-us/product/hd-pro-webcam-c920}, 2013.

\bibitem{Sae-Bae::Gestures::CHI2012}
N.~Sae-Bae, K.~Ahmed, K.~Isbister, and N.~Memon, ``Biometric-rich gestures: A
  novel approach to authentication on multi-touch devices,'' in
  \emph{Proceedings of the 30th ACM SIGCHI Conference on Human Factors in
  Computing Systems (CHI)}, 2012.

\bibitem{Yan::Leakage-Resilient::AsiaCCS2013}
Q.~Yan, J.~Han, Y.~Li, J.~Zhou, and R.~H. Deng, ``Designing leakage-resilient
  password entry on touchscreen mobile devices,'' in \emph{Proceedings of the
  8th ACM Symposium on Information, Computer and Communications Security
  (AsiaCCS)}, 2013.

\bibitem{Kim::authentication::CHI2010}
D.~Kim, P.~Dunphy, P.~Briggs, J.~Hook, J.~W. Nicholson, J.~Nicholson, and
  P.~Olivier, ``Multi-touch authentication on tabletops,'' in \emph{Proceedings
  of the ACM SIGCHI Conference on Human Factors in Computing Systems (CHI)},
  2010.

\bibitem{Biddle::GraphicalPasswords::Surveys2012}
R.~Biddle, S.~Chiasson, and P.~van Oorschot, ``Graphical passwords: Learning
  from the first twelve years,'' in \emph{ACM Computing Surveys}, 2012.

\bibitem{Suo::GraphicalPasswords::ACSAC2005}
X.~Suo, Y.~Zhu, and G.~S. Owen, ``Graphical passwords: A survey,'' in
  \emph{Proceedings of Annual Computer Security Applications Conference
  (ACSAC)}, 2005.

\bibitem{Bulling::Gaze-Based::CHI2012}
A.~Bulling, F.~Alt, and A.~Schmidt, ``Increasing the security of gaze-based
  cued-recall graphical passwords using saliency masks,'' in \emph{Proceedings
  of the ACM SIGCHI Conference on Human Factors in Computing Systems (CHI)},
  2012.

\bibitem{Hirsch::Securekeyboard::1982}
S.~B. Hirsch, ``Secure keyboard input terminal,'' in \emph{United States Patent
  No. 4,333,090}, 1982.

\bibitem{Hirsch::Securekeyboard::1984}
------, ``Secure input system,'' in \emph{United States Patent No. 4,479,112},
  1982.

\bibitem{McIntyre+::securepin::2003}
K.~E. McIntyre, J.~F. Sheets, D.~A.~J. Gougeon, C.~W. Watson, K.~P. Morlang,
  and D.~Faoro, ``Method for secure pin entry on touch screen display,'' in
  \emph{United States Patent No. 6,549,194}, 2003.

\bibitem{Hoanca::Screentechnique::2005}
B.~Hoanca and K.~Mock, ``Screen oriented technique for reducing the incidence
  of shoulder surfing,'' in \emph{Proceedings of the International Conference
  on Security and Management (SAM)}, 2005.

\bibitem{Shin::RandomKeypad::2010}
H.-S. Shin, ``Device and method for inputting password using random keypad,''
  in \emph{United States Patent No. 7,698,563}, 2010.

\bibitem{Lee::RandomKeypad::2011}
C.~Lee, ``System and method for secure data entry,'' in \emph{United States
  Patent Application Publication}, 2011.

\bibitem{Kim::RandomKeypad::2012}
I.~Kim, ``Keypad against brute force attacks on smartphones,'' in \emph{IET
  Information Security}, 2012.

\end{thebibliography}
\end{document}